\documentclass[aps,preprint,prc]{revtex4}
\usepackage{graphics}
\usepackage[colorlinks=true,linkcolor=blue]{hyperref}
\usepackage{latexsym}
\makeatother
\makeatletter
\usepackage[T1]{fontenc}
\makeatletter
\usepackage{latexsym}

\begin{document}
\preprint{AstroUSl-2000-4}
\title{The quark star in the Quark Mean Field approach.}

\author{Ryszard Ma\'{n}ka}
\email{manka@us.edu.pl} \homepage{www.cto.us.edu.pl/~manka}
\thanks{Presented at the workshop \emph{Physics of Neutron Star Interiors} in the European
Center for Theoretical Studies in Nuclear Physics and related
Areas ECT{*} in Trento, Italy, 2000.}
\author{Grzegorz Przyby\l {}a}
\email{przybyla@phys.server.us.edu.pl}
\affiliation{University of
Silesia, Institute of Physics, Katowice 40007, ul. Uniwersytecka
4, Poland}

\date{\today}

\begin{abstract}
The quark star in the quark mean field (QMF) extended to include the strange
mesons interaction is examined. The quark star properties in this model are
studied.
\end{abstract}

\maketitle

\section*{Introduction}

Recent theoretical studies show that the properties of nuclear matter can be
described nicely in terms of the Relativistic Mean Field Theory (RMF) \cite{wal, glen, weber}.
Guichon proposed an interesting model on the change of the nucleon properties
in nuclear matter (QMF or quark-meson coupling model (QMC)) \cite{aust}. The
model construction mimics the relativistic mean field theory, where the scalar
\( \sigma  \) and the vector meson \( \omega  \) fields couple not with nucleons
but directly with quarks. The quark mass has to change from its bare mass due
to the coupling to the \textbf{\( \sigma  \)} meson. In this work we shall
investigate the quark matter within the Relativistic Mean Field theory motivated
by the Nambu-Jona-Lasinio (NJL) model \cite{NJL}\cite{bub}\cite{sch}. The
three-flavor NJL model has been discussed by many authors, e.g. \cite{Rel}.
To include the strange quark the QMF model is extended by incorporating an additional
pair of the strange meson fields which couple only to the \( s \) quark \cite{pal}.
It is very interesting to construct the Guichon model, where the nucleon is
described in terms of constituent quarks, which couple with mesons and gluons.
This model (we refer it as the quark mean field (QMF) model was named in \cite{QMF}).
The meson fields act on quarks inside a nucleon or the neutron or quark star
and change the bulk nuclear properties. The QMF model predicts an increasing
of the nuclear size and a reduction of the nucleon mass in the nuclear environment
of the star.

The properties of strange quark matter (SQM) and the conjecture that SQM could
be the absolute ground state of strongly interacting matter \cite{Bodmer,Witten}
have attracted much attention in nuclear physics and astrophysics. In 1984 Farhi
and Jaffe \cite{Farhi} found that SQM is stable compared with an \( ^{56}Fe \)-nucleus.
Still the bag model calculations play a key role in this field \cite{hea}.

In the framework of the NJL model Babulla and Oertel \cite{bub} have showed
that due to a large constituent quark the strange star is not absolutely stable.
The quark-meson coupling will reduce the effective quark mass what may make
the existence of the strange star more possible.

The aim of this work is to examine the influence of the quark-meson coupling
model on the quark star or neutron star having the quarks core.

\section*{The Quark Mean Field Theory}

The fields of the model QMF for \( \sigma ,\, \omega  \) and \( \rho  \)-mesons
are denoted as \( \varphi  \), \( \omega _{\mu } \), \( \rho _{\mu } \).
To reproduce the observed strongly attractive \( \Lambda \, \Lambda  \) interaction
two additional meson fields are added, the scalar \( f_{0} \) and meson fields
\( \phi  \) denoted as the fields \( \varphi _{*} \)and \( \phi _{\mu } \).
The Lagrange density function for this model has the following form \begin{eqnarray}
{\mathcal{L}}= & -\frac{1}{4}R_{\mu \nu }^{a}R^{a\mu \nu }-\frac{1}{4}F_{\mu \nu }F^{\mu \nu }+\frac{1}{2}M^{2}_{\omega }\, \omega _{\mu }\omega ^{\mu }+\frac{1}{2}M^{2}_{\rho }\, \rho ^{a}_{\mu }\rho ^{a\mu } & \nonumber \\
 & \frac{1}{2}\partial _{\mu }\varphi \partial ^{\mu }\varphi -U(\varphi )+\frac{1}{4}c_{3}(\omega _{\mu }\omega ^{\mu })^{2}+ & \nonumber \\
 & i\sum ^{2}_{f=1}\overline{L_{f}}\gamma ^{\mu }\partial _{\mu }L_{f}-\sum ^{2}_{f=1}g_{f}(\overline{L}_{f}He_{Rf}+h.c.)+ & \label{lag} \\
 & i\overline{q}\gamma ^{\mu }D_{\mu }q-\overline{q}(m_{Q}-g^{q}_{s}\varphi I_{2}-g_{s*}^{q}\varphi _{*}I_{3})q+ & \nonumber \\
 & \frac{1}{2}\partial _{\mu }\varphi _{*}\partial ^{\mu }\varphi _{*}-\frac{1}{2}m^{2}_{s*}\varphi ^{2}_{*}-\frac{1}{4}\phi _{\mu \nu }\phi ^{\mu \nu }+\frac{1}{2}M^{2}_{\phi }\phi _{\mu }\phi ^{\mu }. & \nonumber
\end{eqnarray}
 The field stress tensor of the vector mesons have the following structure:
\begin{equation}
R_{\mu \nu }^{a}=\partial _{\mu }\rho ^{a}_{\nu }-\partial _{\nu }\rho ^{a}_{\mu }+g\varepsilon _{abc}\rho _{\mu }^{b}\rho _{\nu }^{c}
\end{equation}

\begin{equation}
F_{\mu \nu }=\partial _{\mu }\omega _{\nu }-\partial _{\nu }\omega _{\mu }
\end{equation}

\[
G_{\mu \nu }^{a}=\partial _{\mu }W^{a}_{\nu }-\partial _{\nu }W^{a}_{\mu }+gf_{abc}W_{\mu }^{b}W_{\nu }^{c}\]

\begin{equation}
\phi _{\mu \nu }=\partial _{\mu }\phi _{\nu }-\partial _{\nu }\phi _{\mu }
\end{equation}
The covariant derivative acts on baryons as follows:

\begin{equation}
D_{\mu }=\partial _{\mu }+\frac{1}{2}ig^{q}_{\rho }\rho ^{a}_{\mu }\tau ^{a}+ig^{q}_{\omega }\omega _{\mu }I_{2}+ig^{q}_{\phi }\phi _{\mu }I_{3}
\end{equation}
 The potential in RMF \cite{wal, rm2, tm1} is given by \begin{equation}
U(\varphi )=B_{0}+\frac{1}{2}m^{2}_{s}\varphi ^{2}+\frac{1}{3}g_{2}\varphi ^{3}+\frac{1}{4}g_{3}\varphi ^{4}_{s}
\end{equation}
 The fermion fields are composed of quarks and electrons, muons and neutrinos
\begin{eqnarray}
 & q=\left( \begin{array}{c}
u\\
d\\
s
\end{array}\right) ,\, \, I_{2}=\left( \begin{array}{ccc}
1 & 0 & 0\\
0 & 1 & 0\\
0 & 0 & 0
\end{array}\right) \, \, \, I_{3}=\left( \begin{array}{ccc}
0 & 0 & 0\\
0 & 0 & 0\\
0 & 0 & 1
\end{array}\right) \, \, \, \tau ^{a}=\left( \begin{array}{cc}
\sigma ^{a} & 0\\
0 & 0
\end{array}\right)  & \\
 & L_{1}=\left[ \begin{array}{l}
\nu _{e}\\
e^{-}
\end{array}\right] _{L},\, \, \, L_{2}=\left[ \begin{array}{l}
\nu _{\mu }\\
\mu ^{-}
\end{array}\right] _{L},\, \, \, e_{Rf}=(e^{-}_{R},\, \mu ^{-}_{R}). &
\end{eqnarray}
 \( m_{s} \), \( M_{\omega } \), \( M_{\rho } \) are masses assigned to the
mesons fields Here \( q \) denotes a quark field with three flavors, \( u \),
\( d \) and \( s \), and three colors. The Lagrangian function includes also
the nonlinear term \( \frac{1}{4}c_{3}(\omega _{\mu }\omega ^{\mu })^{2} \).
\begin{table}
{\centering \begin{tabular}{|c|c|}
\hline
Parameter&
TM1\cite{rm2, tm1}\\
\hline
\hline
\( M \)&
\( 938\, MeV \)\\
\hline
\( M_{\omega } \)&
\( 783\, MeV \)\\
\hline
\( M_{\rho } \)&
\( 770\, MeV \)\\
\hline
\( m_{s} \)&
\( 511.198\, MeV \)\\
\hline
\( m_{s*} \)&
\( 975\, MeV \)\\
\hline
\( M_{\phi } \)&
\( 1020\, MeV \)\\
\hline
\( m_{Q}=m^{*}_{u}=m^{*}_{d}=m^{*}_{s} \)&
\( 310.6\, MeV \) \cite{bub}\\
\hline
\( m_{Q}=m^{*}_{u}=m^{*}_{d}\neq m^{*}_{s} \)&
\( 367.7\, MeV \),\( 549.5\, MeV \) \cite{sch}\\
\hline
\( g_{3} \)&
\( 7.2325\, fm^{-1} \)\\
\hline
\( g_{4} \)&
\( 0.6183 \)\\
\hline
\( g_{s}=g_{\sigma N}=3g^{q}_{s} \)&
\( 10.0289 \)\\
\hline
\( g_{\omega }=g_{\omega N}=3g^{q}_{\omega } \)&
\( 12.6139 \)\\
\hline
\( g_{\rho }=g_{\rho N}=g^{q}_{\rho } \)&
\( 4.6322 \)\\
\hline
\( g_{s*}=g_{\sigma *\Lambda }=2/3\, g_{s} \) &
 \( g_{s*}=3g^{q}_{s*} \)\\
\hline
\( g_{\phi }=g_{\phi \Lambda }=\sqrt{2}/3\, g_{\omega }=3g^{q}_{\phi } \)&
\( c_{3}=71.3075 \) \\
\hline
\end{tabular}\par}

\caption{\label{tab1}Parameter sets for the Lagrangian function (\ref{lag}).}
\end{table}
 We start with equal dynamical quark masses \( m_{Q}=m^{*}_{u}=m^{*}_{d}=m^{*}_{s}=310.6\, MeV \)
(the \( SU(3) \)-symmetric quark matter with starting \( m_{0,u}=m_{0,d}=m_{0,s}=0 \))\cite{bub}
and a bag pressure \( B_{0}=57.3\textrm{ Me}V/\textrm{fm}^{3} \) originated
from the Nambu-Johna-Lasinio model \begin{equation}
L_{NJL}=i\overline{q}\gamma ^{\mu }D_{\mu }q-\overline{q}m_{0}q+G\sum ^{8}_{a=0}\{(\overline{q}\lambda ^{a}q)^{2}+(i\overline{q}\lambda ^{a}\gamma ^{5}q)^{2}\}.
\end{equation}
We restrict ourselves to the isospin SU(2) symmetric case, \( m_{0}^{u}=m_{0}^{d} \),
so\[
m_{0}=\left( \begin{array}{ccc}
m_{0,u} &  & \\
 & m_{0,d} & \\
 &  & m_{0,s}
\end{array}\right) \]
In the quark massless limit the system has a \( U(3)_{L}\times U(3)_{R} \)
chiral symmetry. The model is not renormalizeable and we have to specify a regularization
scheme for divergent integrals. For simplicity we use a sharp cut-off \( \Lambda  \)
in 3-momentum space. Besides \( \Lambda  \) we have to fix the coupling constants
\( G \) and the current quark masses \( m_{0,u} \) and \( m_{0,s} \). In
most of our calculations we will adopt the parameters of \cite{bub} \( \Lambda  \)
= 602.3 MeV, \( G\Lambda ^{2} \) = 1.835. With \( \Lambda  \), \( G \) and
\( K \) as specified above, chiral symmetry is spontaneously broken. The possible
existence of deconfined quark matter in the interior of neutron stars using
the NJL model was examined in paper \cite{sch}. In the work \cite{sch} the
non-symmetrical quark masses \( m_{Q}=\{m^{*}_{u}=m^{*}_{d}=367.7\, MeV,\, m^{*}_{s}=549.5\, MeV\} \)
are obtained (with starting \( m_{0,u}=m_{0,d} \) = 5.5 MeV, \( m_{0,s} \)=
140.7 MeV).

The Euler equation for \( \Phi _{A}=\{ \)\( \varphi  \), \( \varphi _{*} \), \( \omega _{\mu } \), \( \rho _{\mu } \),
\( \phi _{\mu } \), \( q\} \) fields are \begin{equation}
\label{egg1}
\Box \varphi =m_{s}^{2}\varphi +g_{2}\varphi ^{2}+g_{3}\varphi ^{3}-g^{q}_{s}\overline{q}I_{2}q
\end{equation}
\begin{equation}
\label{eggs5}
\Box \varphi _{*}=m_{s*}^{2}\varphi _{*}-g^{q}_{s}\overline{q}I_{3}q
\end{equation}

\begin{equation}
\label{egg3}
-\partial _{\mu }F^{\mu \nu }=M^{2}_{\omega }\omega ^{\nu }+c_{3}(\omega _{\mu }\omega ^{\mu })\omega ^{\nu }-g^{q}_{\omega }J_{SU(2)}^{\nu }
\end{equation}
\begin{equation}
-\partial _{\mu }\phi ^{\mu \nu }=M^{2}_{\phi }\phi ^{\nu }-g^{q}_{\phi }J^{\nu }_{U(1)}
\end{equation}
 where \begin{equation}
J_{SU(2)}^{\nu }=\overline{q}\gamma ^{\nu }I_{2}q
\end{equation}
 is the SU(2) up and down quark current while \begin{equation}
J_{U(1)}^{\nu }=\overline{q}\gamma ^{\nu }I_{3}q
\end{equation}
is the U(1) strange quark current. The nonabelian gauge field \( \rho ^{\nu ,a} \)
obeys \begin{equation}
\label{egg2}
-D_{\mu }R^{\mu \nu ,a}=M^{2}_{\rho }\, \rho ^{\nu ,a}-g^{q}_{\rho }J_{3}^{\nu }
\end{equation}
 where \begin{equation}
J^{\nu }_{3}=\frac{1}{2}\overline{q}\gamma ^{\nu }\tau ^{3}q
\end{equation}
 is the isospin SU(2) current. In the system we have conservation of {}``nucleon{}''
charge \[
Q_{N}=\int d^{3}xJ^{0}_{SU(2)}\]
 the isospin charge \[
Q_{3}=\int d^{3}xJ^{0}_{3}\, \]
and the {}``hyperon{}'' charge \[
Q_{H}=\int d^{3}xJ^{0}_{U(1)}\, \]
The last is the Dirac equation \begin{equation}
\label{egg4}
i\gamma ^{\mu }D_{\mu }q-(m_{Q}-g^{q}_{s}\varphi I_{2}-g^{q}_{s*}\varphi _{*}I_{3})q=0.
\end{equation}
 The physical system is totally defined by the thermodynamic potential \cite{fet}

\begin{equation}
\Omega =-kTlnTr(e^{-\beta (H-\mu _{N}Q_{N}-\mu _{3}Q_{3}-\mu _{H}Q_{H})})
\end{equation}
 where H is the Hamiltonian of the physical system \begin{equation}
H=\sum _{A}\int d^{3}x\{\partial _{0}\Phi _{A}\pi ^{A}_{\Phi }-\mathcal{L}\}
\end{equation}
 and \( \pi ^{A}=\frac{\partial \mathcal{L}}{\partial (\partial _{0}\Phi _{A})} \)
is a momentum connected to the field \( \Phi _{A} \). The fields \( \Phi _{A}=\{ \)\( \varphi  \), \( \varphi _{*} \), \( \omega _{\mu } \), \( \rho _{\mu } \),
\( \phi _{\mu } \), \( q\} \) denote all fields in the system. In this paper
we shall use the effective potential approach build using the Bogolubov inequality
\cite{rm, rm2}\begin{equation}
\Omega \leq \Omega _{1}=\Omega _{0}(m_{B},\, m_{F})+<H-H_{0}>_{0}
\end{equation}
 \( \Omega _{0} \) is the thermodynamic potential of the trial system as effectively
free quasiparticle system described by the Lagrange function\begin{eqnarray}
 & {\mathcal{L}}_{0}=\frac{1}{2}\partial _{\mu }\overline{\varphi }\partial ^{\mu }\overline{\varphi }-\frac{1}{2}m^{2}_{s}\overline{\varphi }^{2}+\frac{1}{2}\partial _{\mu }\overline{\varphi }_{*}\partial ^{\mu }\overline{\varphi }_{*}-\frac{1}{2}m^{2}_{s}\overline{\varphi }_{*}^{2}+\nonumber  & \\
 & -\frac{1}{4}\overline{R}_{\mu \nu }^{a}\overline{R}^{a\mu \nu }-\frac{1}{4}\overline{F}_{\mu \nu }\overline{F}^{\mu \nu }-\frac{1}{4}\overline{\phi }_{\mu \nu }\overline{\phi }^{\mu \nu } & \\
 & +\frac{1}{2}M_{\omega }^{2}\overline{\omega }_{\mu }\overline{\omega }^{\mu }+\frac{1}{2}M_{\rho }^{2}\overline{\rho }_{\mu }^{a}\overline{\rho }^{a\mu }+\frac{1}{2}M^{2}_{\phi }\overline{\phi }_{\mu }\overline{\phi }^{\mu }+\sum _{f=u,d,s}\overline{q}_{f}(i\gamma ^{\mu }\overline{D}_{\mu }-m_{F,f})q_{f}\nonumber
\end{eqnarray}
Similar to the general case \[
\overline{R}_{\mu \nu }^{a}=\partial _{\mu }\overline{\rho }^{a}_{\nu }-\partial _{\nu }\overline{\rho }^{a}_{\mu }\]
 and \[
\overline{F}_{\omega ,\mu \nu }=\partial _{\mu }\overline{\omega }_{\nu }-\partial _{\nu }\overline{\omega }_{\mu }.\]
 and for gluons \[
\overline{G}_{\mu \nu }^{a}=\partial _{\mu }\overline{W}^{a}_{\nu }-\partial _{\nu }\overline{W}^{a}_{\mu }\]
 We decompose the \( \Phi _{A} \) field into two components, the effectively
free quasiparticle field \( \tilde{\Phi }_{A} \) and the classical boson condensate
\( \xi _{A} \)\begin{equation}
\Phi _{A}=\tilde{\Phi }_{A}+\xi _{A}
\end{equation}
 In the case of the RMF model we have \begin{equation}
\varphi =\overline{\varphi }+\sigma
\end{equation}

\begin{equation}
\varphi _{*}=\overline{\varphi }_{*}+\sigma _{*}
\end{equation}

\begin{equation}
\omega _{\mu }=\overline{\omega }_{\mu }+w_{\mu },\, \, w_{\mu }=\delta _{\mu ,0}w
\end{equation}

\begin{equation}
\rho _{\mu }^{a}=\overline{\rho }^{a}+r^{a}_{\mu },\, \, r^{a}_{\mu }=\delta ^{a,3}\delta _{\mu ,0}\, r
\end{equation}

\begin{equation}
\phi _{\mu }=\overline{\phi }_{\mu }+w_{\mu *},\, \, w_{\mu *}=\delta _{\mu ,0}w_{*}
\end{equation}
The \( \xi _{A}=\{\sigma ,\, w,\, r,\, w_{*}\} \) field will be treated as
the variational parameters in the effective potential. Also the boson and fermion
mass \( m_{B},\, m_{F} \) will be treated as as the variational parameters.
The covariant derivative for the trial system is \begin{equation}
\overline{D}_{\mu }=\partial _{\mu }+\frac{1}{2}ig^{q}_{\rho }r^{a}_{\mu }\sigma ^{a}+ig^{q}_{\omega }w_{\mu }+ig^{q}_{\phi }w_{\mu *}
\end{equation}
 This introduce the homogenous fermion interaction with boson condensate \( w_{\mu } \),
\( r^{a}_{\mu } \), \( w_{\mu *} \). The fermion quasiparticle will obey the
Dirac equation \begin{equation}
(i\gamma ^{\mu }\, \overline{D}_{\mu }-m_{F})\psi _{f}=0
\end{equation}
 The constant condensate \( w,\, r \) simply shift the chemical potential from
\( \mu _{i}=\mu  \)\( ^{0}_{i} \) (when \( w=r=0 \)) to \begin{eqnarray}
\mu _{u}=\mu ^{0}_{u}+\frac{1}{2}g^{q}_{\rho }r-g^{q}_{\omega }w &  & \label{mun} \\
\mu _{d}=\mu ^{0}_{d}-\frac{1}{2}g^{q}_{\rho }r-g^{q}_{\omega }w &  & \label{mup} \\
\mu _{s}=\mu ^{0}_{s}-g^{q}_{\phi }w_{*} &  &
\end{eqnarray}
 where \( \mu _{u}=\mu +\frac{1}{2}\mu _{3} \) and \( \mu _{d}=\mu -\frac{1}{2}\mu _{3} \).\\
 Quarks and electrons are in \( \beta  \)-equilibrium which can be described
as a relation among their chemical potentials \[
\mu _{d}=\mu _{u}+\mu _{e}=\mu _{s}\]
 where \( \mu _{u} \), \( \mu _{d} \), \( \mu _{s} \) and \( \mu _{e} \)
stand for quarks and electron chemical potentials respectively. If the electron
Fermi energy is high enough (greater then the muon mass) in the neutron star
matter muons start to appear as a result of the following reaction \begin{eqnarray*}
d\rightarrow u+e^{-}+\overline{\nu }_{e} &  & \\
s\rightarrow u+\mu ^{-}+\overline{\nu }_{\mu } &  &
\end{eqnarray*}
 In a pure quark state the star should to be charge neutral. This gives us an
additional constraint on the chemical potentials \begin{equation}
\frac{2}{3}n_{u}-\frac{1}{3}n_{d}-\frac{1}{3}n_{s}-n_{e}=0.
\end{equation}
 where \( n_{f} \) (\( f\in u,d,s \)), \( n_{e} \) the particle densities
of the quarks and the electrons, respectively. The EOS can now be parameterized
by only one chemical potential, say \( \mu _{u}=M\, x \). Variation

\[
\frac{\partial f_{1}}{\partial m_{F}}=0\]
 with respect to the trial system \( L_{0} \) gives \begin{eqnarray}
m_{F,u,d}=m_{Q}-g^{q}_{s}\sigma =M\delta _{u,d} &  & \\
m_{F,s}=m_{Q}-g^{q}_{s*}\sigma =M\delta _{s} &  &
\end{eqnarray}
 In the local equilibrium inside the star the free energy reaches the minimum
at \( \sigma  \).

The same result may be achieved calculating the averages of the equation of
motions (\ref{egg1}) for the effective system \( {\mathcal{L}}_{0} \). In
the mean field approximation the meson field operators are replaced by their
expectation values. We also consider the isotropic system at rest. As \( <\overline{q}q>_{0} \)
(calculated with respect to \( {\mathcal{L}}_{0} \) system) depends on the
effective quark mass \( m_{F} \) (or \( \sigma ,\, \sigma _{*} \) ) the equation
(\ref{egg1}) is highly nonlinear also with respect to \( \sigma  \).

Calculation base on the relation \[
\frac{\partial f_{F}}{\partial m_{F,f}}=<\overline{q}q>_{0}\]
 gives \begin{eqnarray}
<\overline{q}_{f}q_{f}>_{0}=\frac{m_{F}}{\pi ^{2}}\int _{0}^{\infty }\frac{p^{2}dp}{\sqrt{p^{2}+m_{F}^{2}}}\{\frac{1}{\exp (\beta (\epsilon _{p}-\mu _{f}))+1}+ &  & \\
\frac{1}{\exp (\beta (\epsilon _{p}+\mu _{f}))+1}\}. &  & \nonumber
\end{eqnarray}
 The quantum average \( <\overline{q}_{f}q_{f}>_{0} \) depends on the quark
chemical potentials (\ref{mun},\ref{mup}). In the result the effective quark
effective mass \( m_{F} \) also will be dependent on the chemical potentials.
In the result the solution \( \sigma  \) of the equation (\ref{egg1}) also
will be dependent. The same situation will consider other fields.
\begin{figure}
{\par\centering \resizebox*{12cm}{!}{\includegraphics{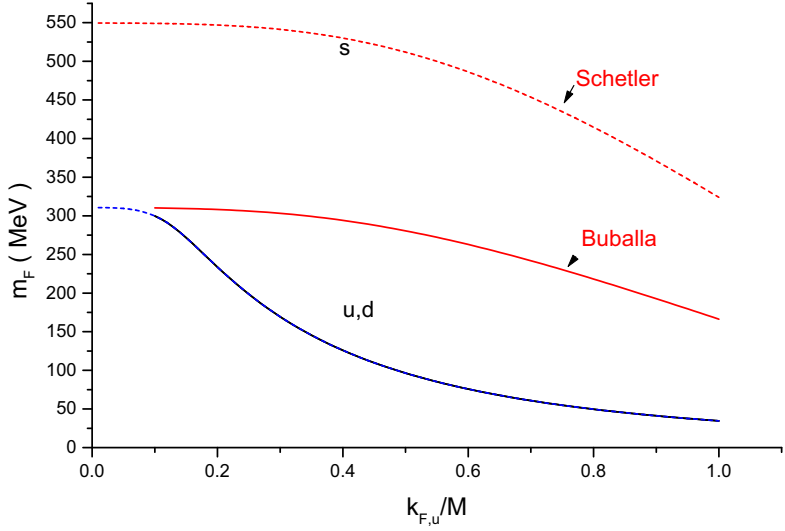}} \par}

\caption{\label{figd} The effective quark masses.}
\end{figure}

\begin{figure}
{\par\centering \resizebox*{12cm}{!}{\includegraphics{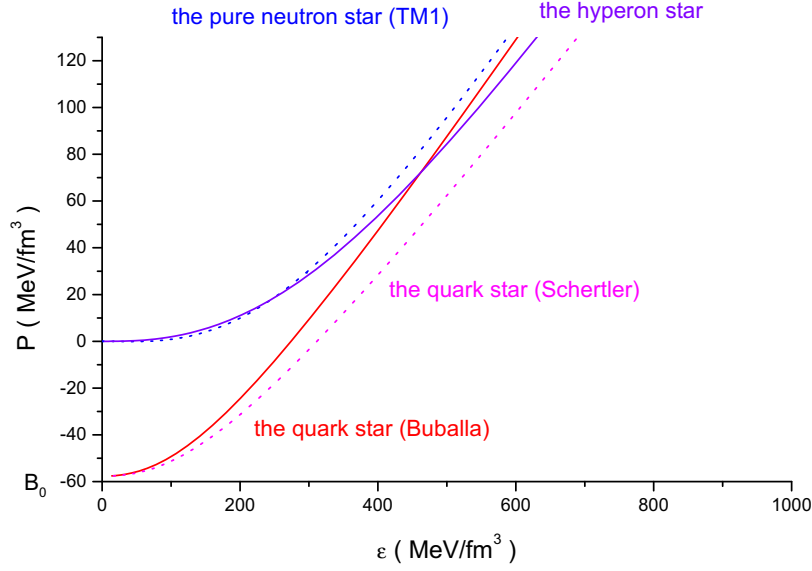}} \par}

\caption{\label{feso} The quark and neutron (TM1) and hyperon equation of state.}
\end{figure}

The effective quarks mass \( m_{F,f} \) (or \( \delta _{f}=m_{F,f}/M \) )
dependence on the dimensionless Fermi momentum \( x_{F} \) is presented on
the Fig.\ref{figd}. To calculate the properties of the quark star we need the
energy-momentum tensor. We define the density of energy and pressure by the
energy - momentum tensor \[
<T_{\mu \nu }>=(P+\epsilon )u_{\mu }u_{\nu }-Pg_{\mu \nu }\]
 where \( u_{\mu } \) is a unite vector (\( u_{\mu }u^{\mu }=1 \)). Similar
to paper \cite{rm2, toki} we have introduced the dimensionless {}``Fermi{}''
momentum even at finite temperature which exactly corresponds to the Fermi momentum
at zero temperature. Both \( \epsilon _{F} \) and \( P_{F} \) depend on the
quark chemical potential \( \mu  \) or Fermi momentum \( x_{F} \). This parametric
dependence on \( \mu  \) (or \( x_{F} \)) defines the equation of state. The
various equations of state for different parameters sets is presented on Fig.\ref{feso}.

\section*{The quark star properties}

\begin{figure}
{\par\centering \resizebox*{12cm}{!}{\includegraphics{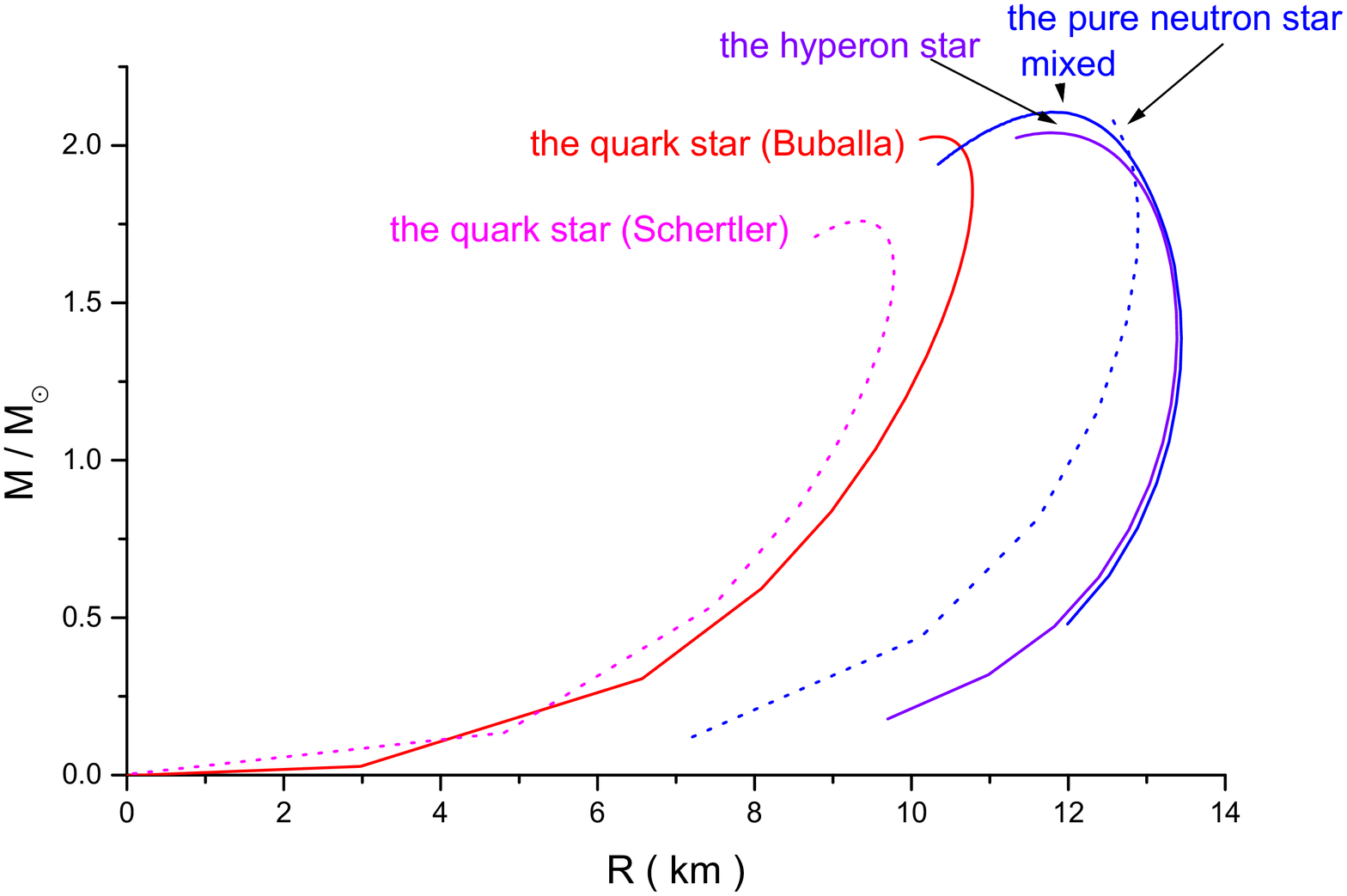}} \par}

\caption{\label{figrm} The \protect\( M(R)\protect \) dependence for the quark star.}
\end{figure}

\begin{figure}
{\par\centering \resizebox*{12cm}{!}{\includegraphics{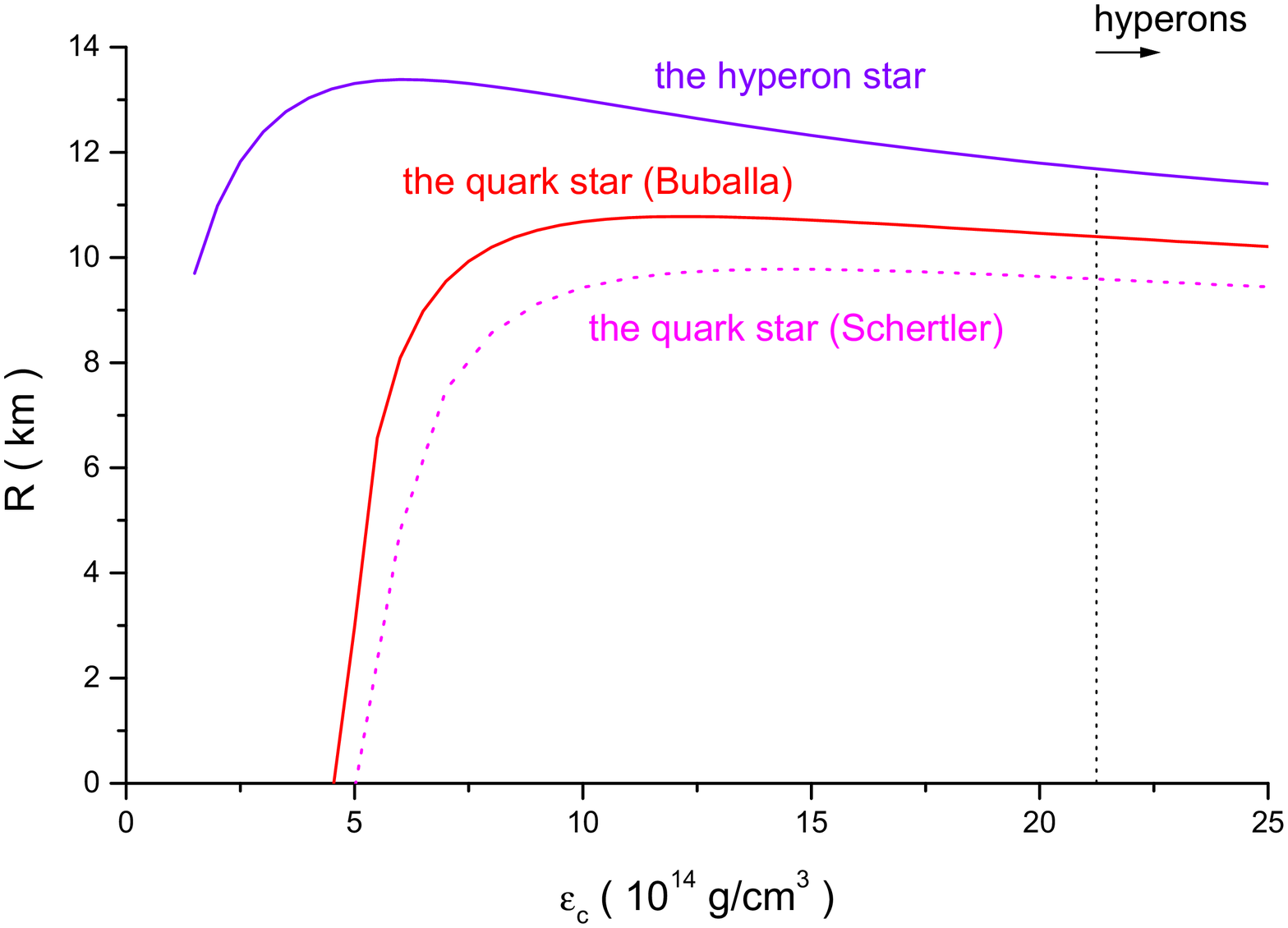}} \par}

\caption{\label{figrho} The \protect\( R(\rho )\protect \) dependence for the quark
star.}
\end{figure}

\begin{figure}
{\par\centering \resizebox*{12cm}{!}{\includegraphics{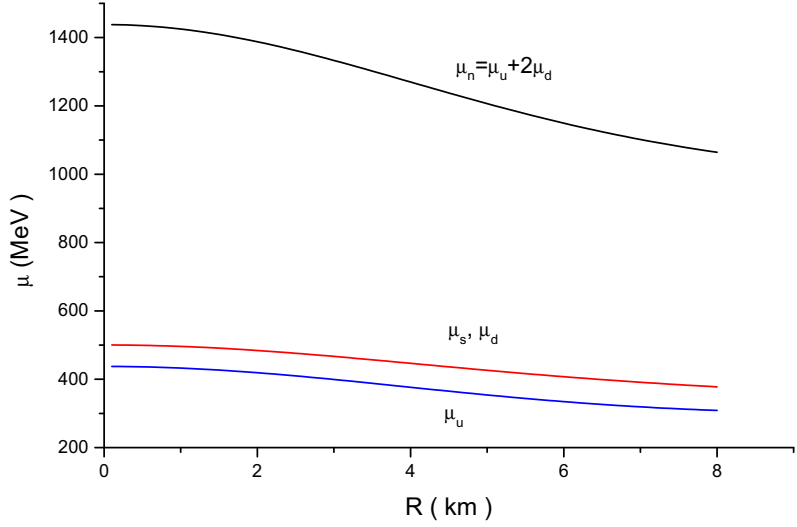}} \par}

\caption{\label{figmu}The effective quark chemical potential for the quark star with
the central density \protect\( \rho _{c}=2\, 10^{15}\, g/cm^{3}\protect \)}
\end{figure}

\begin{figure}
{\par\centering \resizebox*{12cm}{!}{\includegraphics{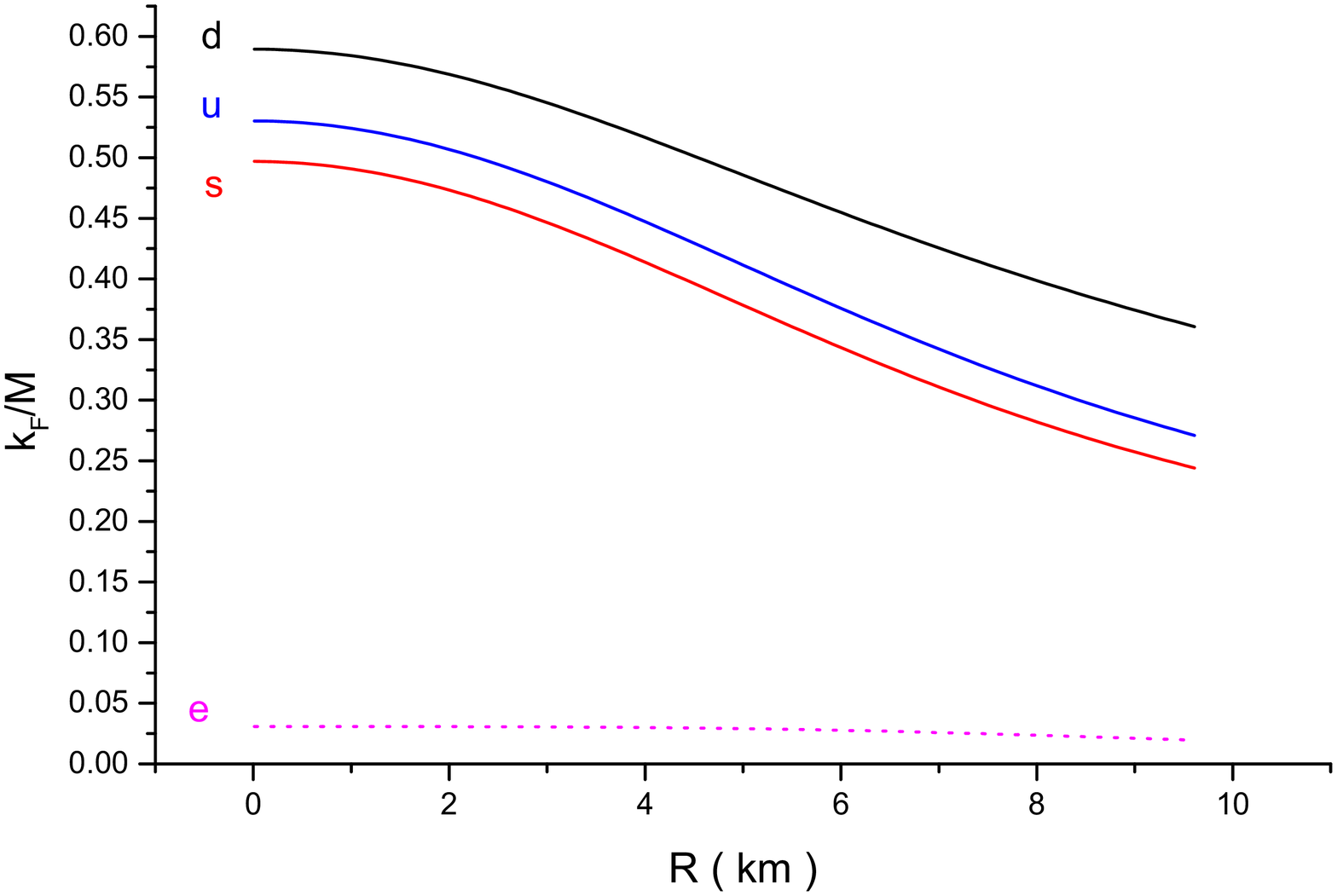}} \par}

\caption{\label{figuds}The quark and electron Fermi momentum inside the star }
\end{figure}

The numerical results describing the structure of neutron or quark star are
based on the relativistic mean field theory are now presented. It is possible
to describe a static spherical star solving the OTV equation. \begin{equation}
\label{teq1}
\frac{dP(r)}{dr}=-\frac{G}{r^{2}}(\rho (r)+\frac{P(r)}{c^{2}})\frac{(m(r)+\frac{4\pi }{c^{2}}P(r)r^{3})}{(1-\frac{2Gm(r)}{c^{2}r})}
\end{equation}
\begin{equation}
\label{teq2}
\frac{dm(r)}{dr}=4\pi r^{2}\rho (r)
\end{equation}
 Having solved the OTV equation the pressure \( p(r) \), mass \( m(r) \) and
density \( \rho (r) \) were obtained. To obtain the total radius \( R \) of
the star the fulfillment of the condition \( p(R)=0 \) is necessary. This allows
to determine the total gravitational mass of the star \( M(R) \). The \( M(R) \)
for the quark star is presented on the Fig. \ref{figrm}.
 The \( R(\rho ) \) dependence for the quark star is presented on the Fig.
\ref{figrho}.
 For the quark star with the central density \( \rho _{c}=2\, 10^{15}\, g/cm^{3} \)
the star profile in the mean field approach is presented on the Figs. \ref{figmu},\ref{figuds},\ref{figw}.
The effective quark chemical potential are presented on the Fig. \ref{figmu}.

\begin{figure}
{\par\centering \resizebox*{12cm}{!}{\includegraphics{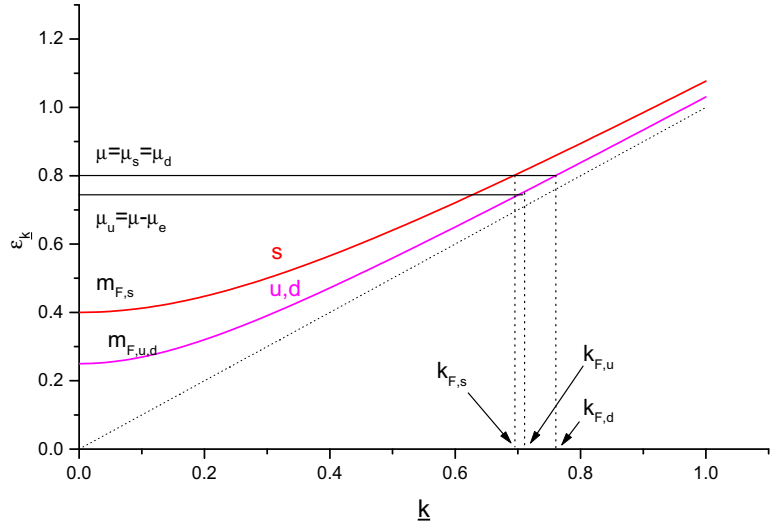}} \par}

\caption{\label{Figbet}The quark Fermi momentum distribution.}
\end{figure}

\begin{figure}
{\par\centering \resizebox*{12cm}{!}{\includegraphics{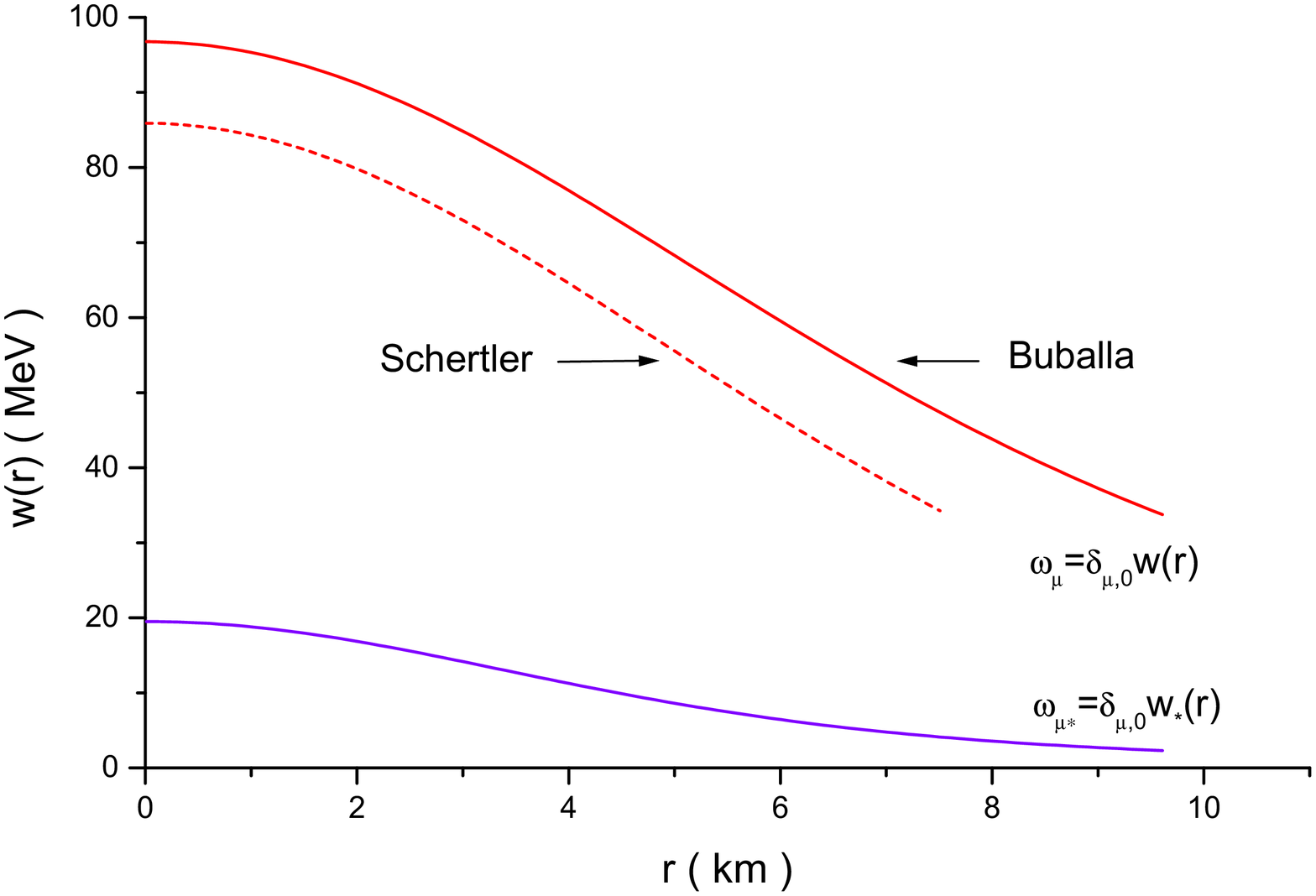}} \par}

\caption{\label{figw}The meson field \protect\( \omega _{\mu }\protect \) and \protect\( \phi _{\mu }\protect \)
profile inside the quark star.}
\end{figure}

\begin{figure}
{\par\centering \resizebox*{12cm}{!}{\includegraphics{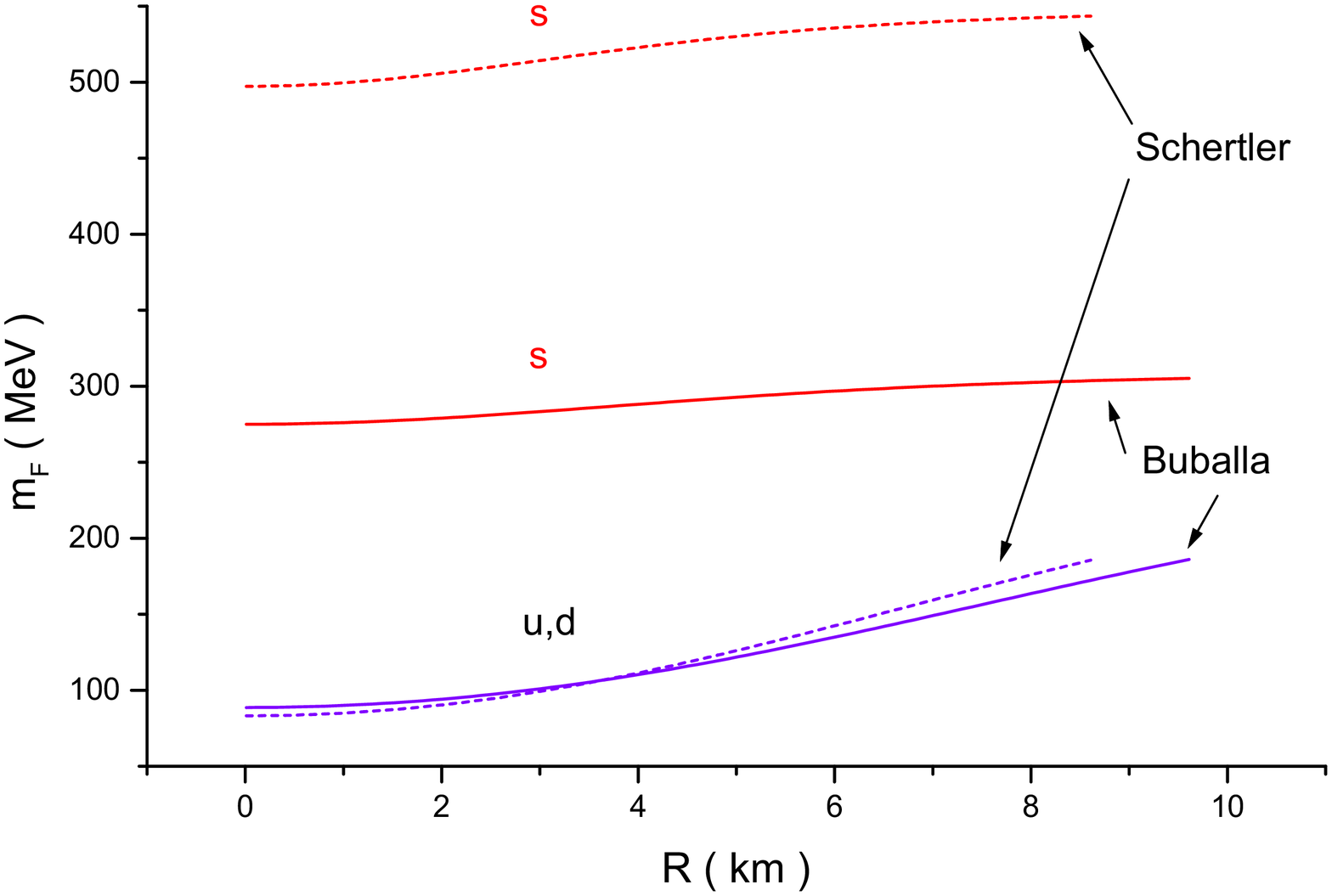}} \par}

\caption{\label{Figdel}The quark effective mass profile inside the star.}
\end{figure}

\begin{figure}
{\par\centering \resizebox*{12cm}{!}{\includegraphics{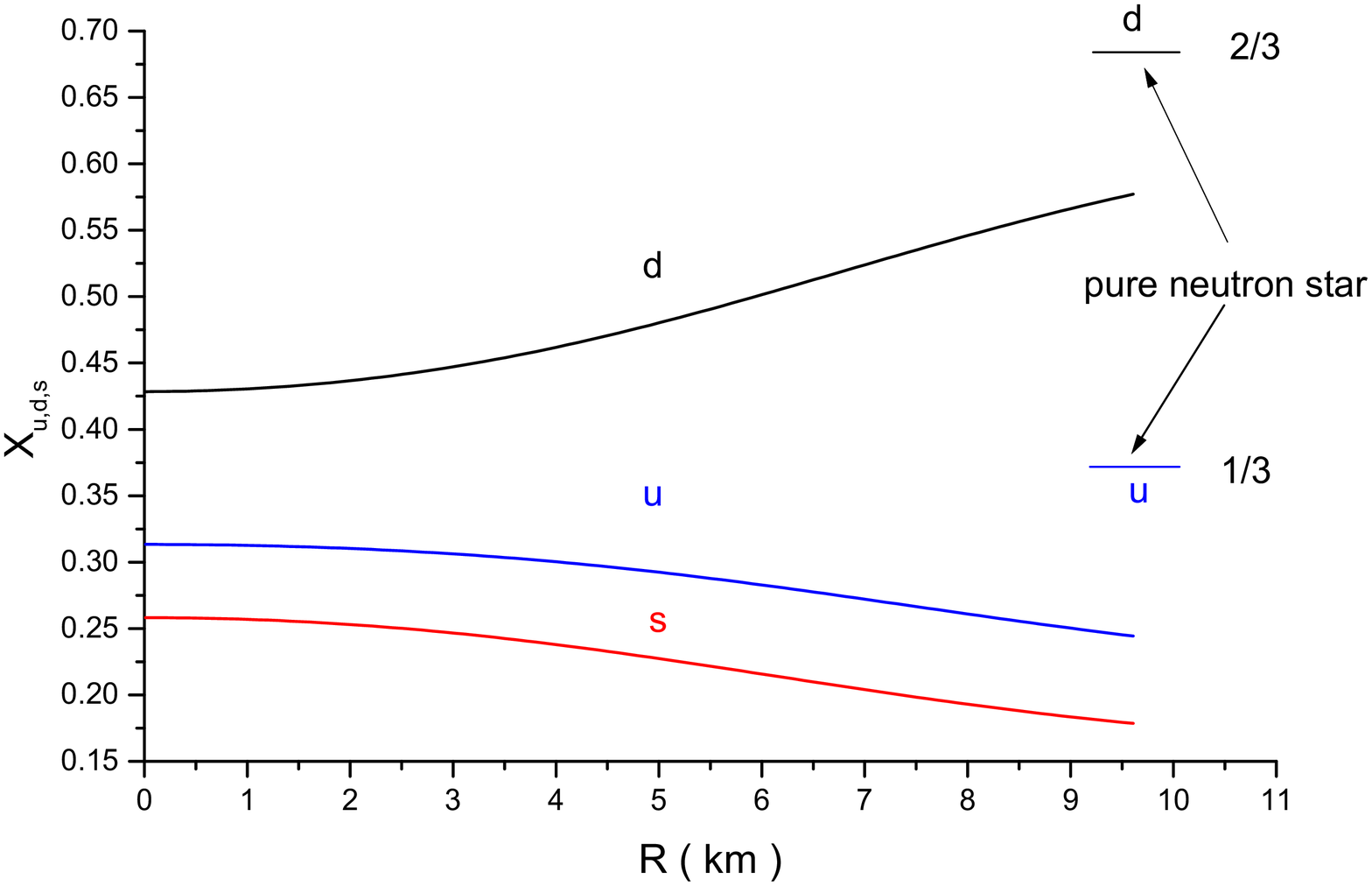}} \par}

\caption{\label{Xuds}The quark partial fraction inside the star.}
\end{figure}

\begin{figure}
{\par\centering \resizebox*{12cm}{!}{\includegraphics{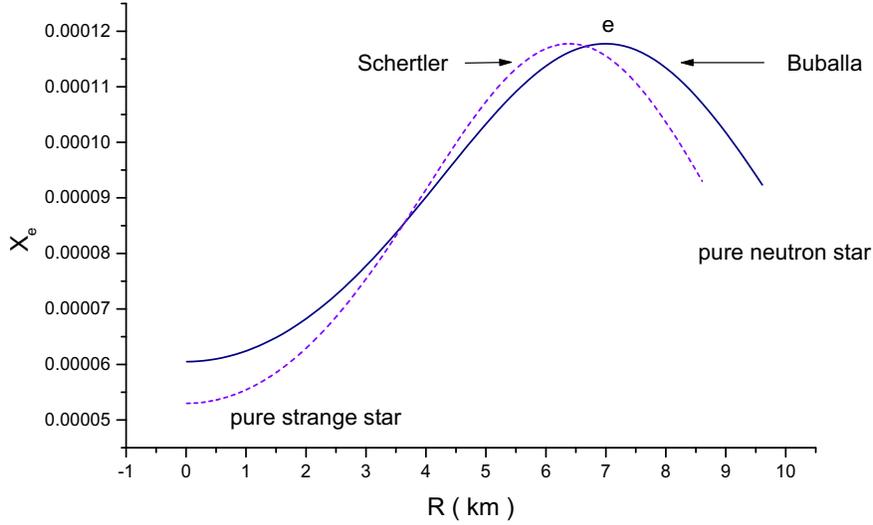}} \par}

\caption{\label{Fige}The electron distribution inside the quark star.}
\end{figure}

The quark and electron Fermi momentum inside the star is presented on Fig. \ref{figuds}.
This Fermi momentum distribution comes from the quark dispersion relation presented
on the Fig. \ref{Figbet}.
 The meson field \( \omega _{\mu } \) and \( \phi _{\mu } \) profile inside
the star is presented on the Fig. \ref{figw}.
The biggest contributions to the energy density come from the quarks, the gauge
boson field \( \omega _{\mu } \), \( \phi _{\mu } \) and scalar boson field
\( \varphi  \), \( \varphi _{*} \). The quark effective mass profile inside
the star is presented on Fig. \ref{Figdel}. This difference between strange
and up, down quarks deepens inside the star.
 Its influences on the quark partial fraction defined as \begin{equation}
X_{f}=\frac{n_{f}}{(n_{u}+n_{d}+n_{s})},
\end{equation}
where \( f=(u,d,s) \), presented on the Fig. \ref{Xuds}.
We see that both pure neutron star \( (X_{d}=2/3,\, X_{u}=1/3) \) and strange
star \( (X_{u}=X_{d}=X_{s}) \) are rather mathematical limits. The electron
distribution inside the quark star in presented on the Fig. \ref{Fige}.

In the flat spacetime the preferred configuration is realized for the system
with minimum of free energy. As pressure is defined as \[
P=-\frac{\partial F}{\partial V},\]
this means that the stable configuration should have bigger pressure. According
to the equation of state (Fig. \ref{feso}) as was pointed in the strange star
configuration is unstable. As a consequence there seems to be no chance to find
SQM with energies per baryon number lower than in ordinary nuclei. This almost
rules out the original idea of absolutely stable SQM.

However, when hyperon are included there is crossing point at \[
\rho _{cs}=8.92\, 10^{14}\, g/cm^{3}\]
above which the strange quark core (the Buballa, Oertel parameters set \cite{bub})
may appear. This will happen when the star central density \( \rho _{c} \)
exceed \( \rho _{cs} \). However, Fig.\ref{figrho} shows that the hyperon
star is unstable in this region. A mixed hyperon star with strange quark core
can be stabilized if the the equation of state will be more softer (including
the kaon condensation \cite{kaon}, for example).

\begin{figure}
{\par\centering \resizebox*{10cm}{!}{\includegraphics{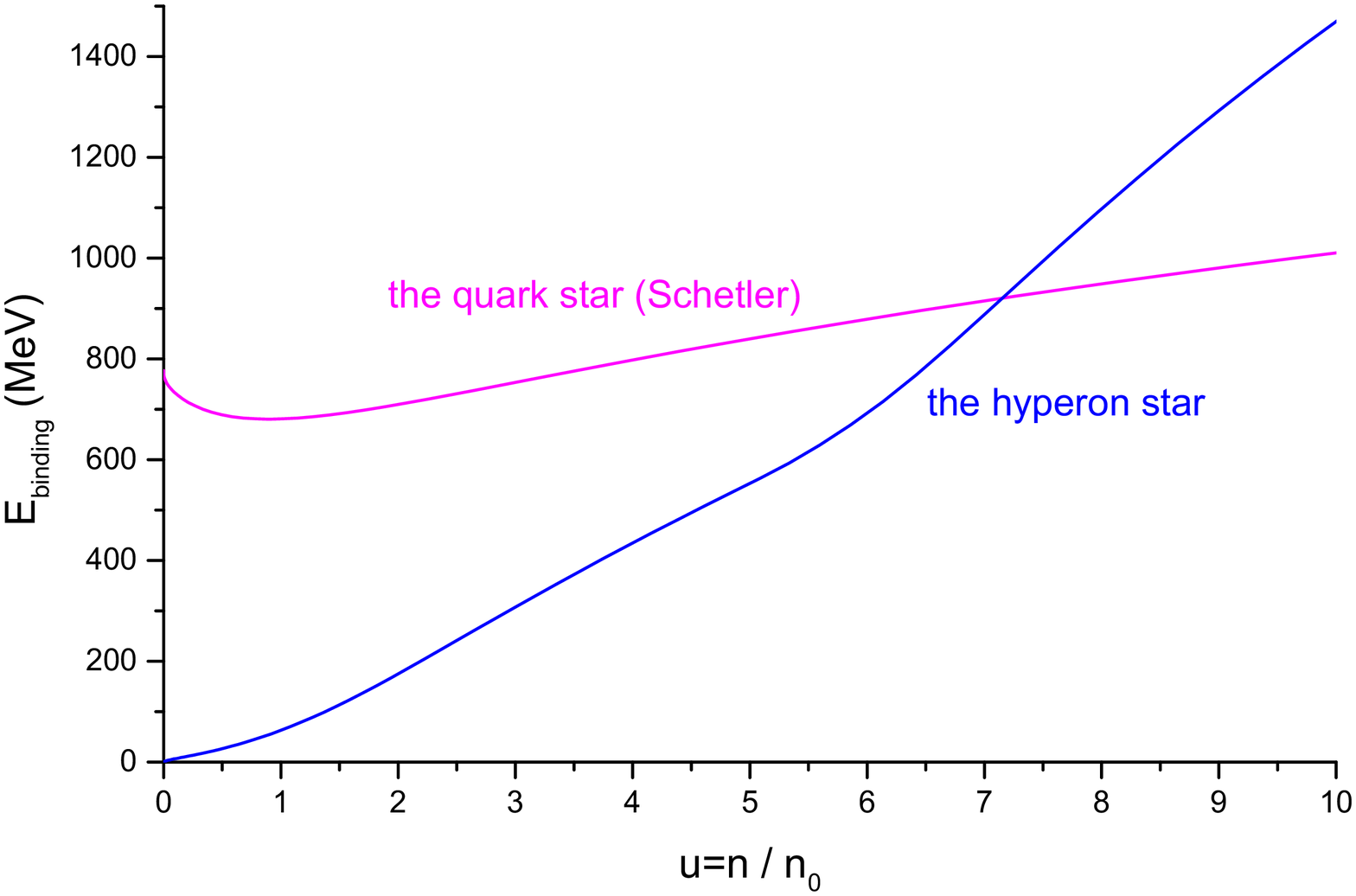}} \par}

\caption{\label{figb}The nuclear binding energy as the relative baryon number \protect\( u=n/n_{0}\protect \)
(\protect\( n_{0}=0.17\, fm^{-3}\protect \) for the nuclear symmetric matter).}
\end{figure}

Unfortunately, the strange star with Schetler at al. parameters set \cite{sch}
is unstable due to a large constituent strange quark mass (\( m^{*}_{s}=549.5\, MeV \)).
In the flat Minkowski spacetime the nucleon binding energy is presented on Fig.\ref{figb}.

\begin{figure}
{\par\centering \resizebox*{10cm}{!}{\includegraphics{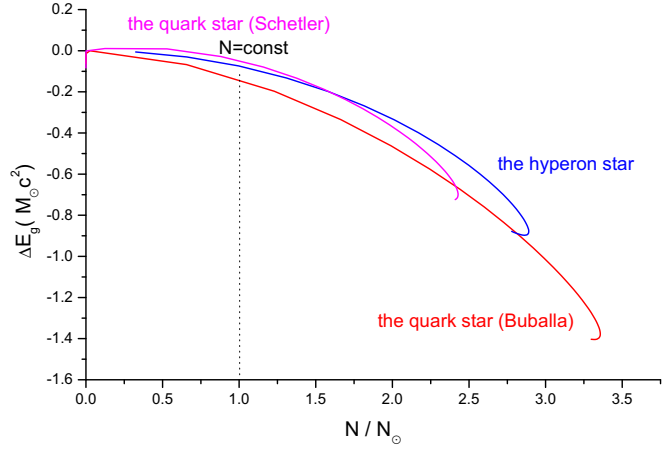}} \par}

\caption{\label{figbin}The star gravitational binding energy as a function of the star
baryon number \protect\( N/N_{\odot }\protect \) (\protect\( N_{\odot }=1.187\, 10^{57}\protect \)
is the baryon number of the Sun).}
\end{figure}

\begin{figure}
{\par\centering \resizebox*{10cm}{!}{\includegraphics{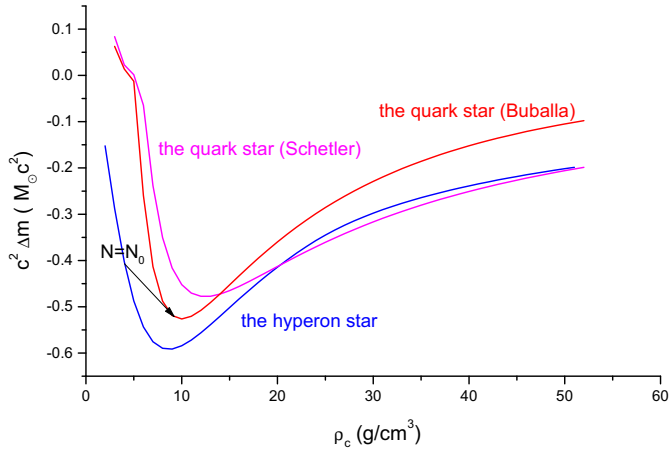}} \par}

\caption{\label{figde}The star mass deficit calculated in the Sun units (\protect\( M_{\odot }c^{2}=1.787\, 10^{54}\, \, ergs\protect \))
as a function of the star central density.}
\end{figure}

All these considerations concerning the star stability ware done in the flat
Minkowski spacetime. However, a neutron or quark star is mainly binded by gravity.
They binding energy \cite{wein} are equal to \begin{equation}
\Delta E=c^{2}M-c^{2}M_{B}=(c^{2}M-c^{2}M_{p})+(c^{2}M_{p}-c^{2}M_{B}),
\end{equation}
where\begin{equation}
M_{p}=\int _{0}^{R}4\pi r^{2}dr(1-\frac{2Gm(r)}{c^{2}r})^{-\frac{1}{2}}\rho (r)
\end{equation}
is the proper mass of the compact object and\begin{equation}
M=\int _{0}^{R}4\pi r^{2}dr\rho (r)
\end{equation}
is the gravitational mass. The gravitational binding energy for the hyperon
and quark star is presented on Fig. \ref{figbin}.
\( M_{B} \) is the baryonic mass defined as\begin{equation}
M_{B}=\sum _{B}m_{B}N_{B}
\end{equation}
where \( m_{B} \) is the baryon mass and the baryon number is equal to \begin{equation}
N_{B}=\int _{0}^{R}4\pi r^{2}dr(1-\frac{2Gm(r)}{c^{2}r})^{-\frac{1}{2}}n_{B}(r)
\end{equation}
The star mass deficit\[
\Delta M=(M_{p}-M)\]
is presented on Fig.\ref{figde}.
The quark star is stable with respect to the gravitational binding energy (Fig.
\ref{figbin}). The difference of the gravitational binding energy for the hyperon
and quark star is order of \begin{equation}
\sim 8.9\, 10^{52}\, \, ergs
\end{equation}
Such a doze of energy may be released during conversion of the neutron star
into quark star. For example, the \( M_{B}=M_{\odot } \) hyperon star with
the mass \( M\sim 0.92\, M_{\odot } \), radius \( R\sim 13\, km \) and central
density \( \rho _{c}\sim 4\, g/cm^{3} \) will transform into the quark star
with the mass \( M\sim 0.85\, M_{\odot } \), radius \( R\sim 8.5\, km \) and
central density \( \rho _{c}\sim 8.9\, g/cm^{3} \) (Figs. \ref{figbin},\ref{figde}).
This conversion may explain nature of the gamma-ray bursts \cite{ign}.

\section*{Conclusion}

Quark meson coupling model was designed to describe both the bulk and internal
structures of nuclear phenomenology described in terms of the Relativistic Mean
Field Theory (RMF) \cite{wal}. The hyperon-hyperon interaction mediated by
the \( \varphi _{*} \)and \( \phi _{\mu } \) mesons is expected to be important
for hyperon rich matter present in the cores of neutron star or core of the
quark star. The quarks interaction is a bit different for the \( u,\, d \)
and \( s \) quarks. This interaction breaks the starting SU(3) flaver symmetry.
The real quark star still is dominated by the \( u,\, d \) quarks with smaller
presence of the \( s \) quarks. Only in the mathematical limit of the core
high density we have the strange star \( (X_{u}=X_{d}=X_{s}) \) case. The appearance
of deconfined quark matter in the center of the neutron star demands soft nuclear
equation of state. The quark star occurs stable when the gravitational binding
takes into consideration.

\section*{Appendix: The hyperon star in the Relativistic Mean Field Theory}

To compare the quark star with neutron one we shall use the RMF model of the
neutron star including the \( \Lambda  \) hyperon \cite{hyp}. This is a simple
generalization of the paper \cite{rm2}. The model consists following mesons:
\( \varphi  \), \( \omega _{\mu } \), \( \rho ^{a}_{\mu } \) with masses
\( m_{s}=510\, MeV \), \( M_{\omega }=783\, MeV \), \( M_{\rho }=770\, MeV \),
and \( \varphi _{*} \), \( \phi _{\mu } \) with masses \( m_{s*}=975\, MeV \),
\( M_{\phi }=1020\, MeV \). The fermion fields are composed of protons, neutrons
and \( \Lambda  \) hyperon\[
\psi =\left( \begin{array}{c}
\psi _{p}\\
\psi _{n}
\end{array}\right) ,\, \, \, \psi _{\Lambda }\]
with masses \( M=938\, MeV \), \( M_{\Lambda }=1115.6\, MeV \).

\begin{figure}
{\par\centering \resizebox*{13cm}{!}{\includegraphics{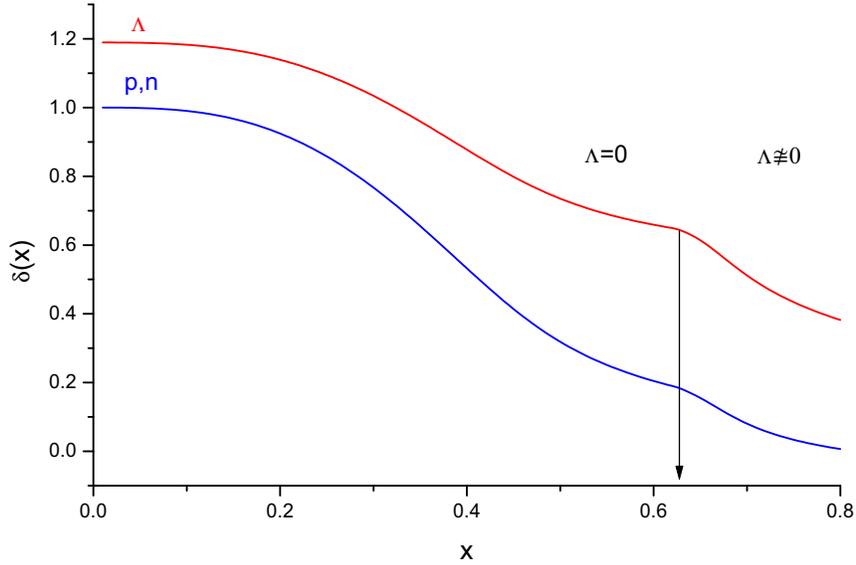}} \par}

\caption{\label{fig1}The effective mass \protect\( \delta =m^{*}_{F}/M\protect \)
for nucleon and \protect\( \Lambda \protect \) as function of the nucleon Fermi
momentum \protect\( x=k_{F}/M\protect \). }
\end{figure}

The Lagrange function is:

\begin{eqnarray}
L=-\frac{1}{2}\partial _{\mu }\varphi \partial ^{\mu }\varphi -U(\varphi )-\frac{1}{4}F_{\mu \nu }F^{\mu \nu }-\frac{1}{2}M_{\omega }^{2}\omega _{\mu }\omega ^{\mu } &  & \\
-\frac{1}{4}c_{3}(\omega _{\mu }\omega ^{\mu })^{2}-\frac{1}{4}R_{\mu \nu }^{a}R^{a\mu \nu }-\frac{1}{2}M_{\rho }^{2}\rho _{\mu }^{a}\rho ^{a\mu } &  & \\
-\frac{1}{2}\partial _{\mu }\varphi _{*}\partial ^{\mu }\varphi _{*}-\frac{1}{2}m_{s*}^{2}\varphi ^{2}_{*}-\frac{1}{4}\Phi _{\mu \nu }\Phi ^{\mu \nu }-\frac{1}{2}M_{\phi }^{2}\phi _{\mu }\phi ^{\mu } &  & \\
+i\overline{\psi }\gamma ^{\mu }D_{\mu }\psi -\overline{\psi }(M-g_{\sigma N}\varphi )\psi  &  & \\
+i\overline{\psi }_{\Lambda }\gamma ^{\mu }D_{\mu }\psi _{\Lambda }-\overline{\psi }_{\Lambda }(M_{\Lambda }-g_{\sigma \Lambda }\varphi -g_{\sigma _{*}\Lambda }\varphi _{*})\psi _{\Lambda } &  &
\end{eqnarray}
where \( F_{\mu \nu } \) is the stress tensor of the form \begin{eqnarray*}
F_{\mu \nu }=\partial _{\mu }\omega _{\nu }-\partial _{\nu }\omega _{\mu },
\end{eqnarray*}
 and the tensor \( R_{\mu \nu }^{a} \) has the form \begin{eqnarray*}
R_{\mu \nu }^{a}=\partial _{\mu }\rho _{\nu }^{a}-\partial _{\nu }\rho _{\mu }^{a}+g_{\rho }\varepsilon _{abc}\rho _{\mu }^{b}\rho _{\nu }^{c}
\end{eqnarray*}
 and the scalar potential is \begin{eqnarray*}
U(\varphi )=\frac{1}{2}M_{\sigma }^{2}\varphi ^{2}+\frac{1}{3}g_{2}\varphi ^{3}+\frac{1}{4}g_{3}\varphi ^{4},
\end{eqnarray*}
 where \( c_{3} \), \( g_{2} \), \( g_{3} \) are constant of the TM1 model
\cite{rm, rm2}. For the hyperon sector \begin{eqnarray*}
\Phi _{\mu \nu }=\partial _{\mu }\phi _{\nu }-\partial _{\nu }\phi _{\mu }, &
\end{eqnarray*}
The covariant derivative are defined now as \begin{eqnarray}
D_{\mu }\psi =(\partial _{\mu }+\frac{1}{2}ig_{\rho N}\rho _{\mu }^{a}\sigma ^{a}+ig_{\omega N}\omega _{\mu })\psi , &  & \\
D_{\mu }\psi _{\Lambda }=(\partial _{\mu }+ig_{\phi \Lambda }\phi _{\mu })\psi _{\Lambda } &  &
\end{eqnarray}
The coupling constants are\begin{eqnarray*}
g_{\sigma \Lambda }=\frac{2}{3}g_{\sigma N} &  & \\
g_{\sigma *\Lambda }=\frac{\sqrt[]{2}}{3}\, g_{\sigma N} &  &
\end{eqnarray*}
for SU(6) symmetry, the coupling constants \( g_{\sigma N} \) , \(
g_{\omega N} \) are the same as for the TM1 parameters set
\cite{rm2} and \[ g_{\phi \Lambda }=\frac{\sqrt[]{2}}{3}\,
g_{\omega N}\] \cite{pal}. The effective mass of nucleon
is\begin{equation} m_{N}^{*}=M\, \delta =M-g_{\sigma \Lambda
}\sigma
\end{equation}
and \begin{equation}
m^{*}_{\Lambda }=M_{\Lambda }=M\, \delta _{\Lambda }=g_{\sigma \Lambda }\sigma -g_{\sigma _{*}\Lambda }\sigma _{*}
\end{equation}
for the \( \Lambda  \) hyperon. The effective mass is presented on Fig \ref{fig1}.
This model allows us to calculate the equation of state presented on Fig.\ref{feso}.

\end{document}